\documentclass[letter]{emulateapj}
\pdfoutput=1
\usepackage{graphicx}
\usepackage[large]{subfigure}
\usepackage{amssymb, amsmath}
\usepackage[amssymb]{SIunits}
\usepackage{tabularx}
\usepackage{mathrsfs}
\usepackage{color}
\usepackage{float}
\usepackage{amsmath}
\usepackage{natbib}
\usepackage[breaklinks, colorlinks, citecolor=blue]{hyperref}
\newcommand{\be}{\begin{equation}}
\newcommand{\ee}{\end{equation}}
\newcommand{\ba}{\begin{eqnarray}}
\newcommand{\ea}{\end{eqnarray}}
\newcommand{\delz}{\Delta z_{\rm re}}
\newcommand{\zre}{z_{\rm re}}
\newcommand{\xe}{x_{\rm i}}

\shorttitle{Reionization from next-generation CMB experiments}
\shortauthors{Calabrese, Hlo\v{z}ek, Battaglia et al.}

\begin{document}

\title{Precision Epoch of Reionization studies with next-generation CMB experiments}
\author{Erminia~Calabrese\altaffilmark{1}, 
Ren\'{e}e~Hlo\v{z}ek\altaffilmark{2}, 
Nick~Battaglia\altaffilmark{3}, 
J.~Richard~Bond\altaffilmark{4},
Francesco~de~Bernardis\altaffilmark{5},
Mark~J.~Devlin\altaffilmark{6},
Amir~Hajian\altaffilmark{4},
Shawn~Henderson\altaffilmark{5},
J.~Colin~Hill\altaffilmark{2},
Arthur~Kosowsky\altaffilmark{7},
Thibaut~Louis\altaffilmark{1}, 
Jeff~McMahon\altaffilmark{8},
Kavilan~Moodley\altaffilmark{9},
Laura~Newburgh\altaffilmark{10},
Michael~D.~Niemack\altaffilmark{5},
Lyman~A.~Page\altaffilmark{11},
Bruce~Partridge\altaffilmark{12},
Neelima~Sehgal\altaffilmark{13},
Jonathan~L.~Sievers\altaffilmark{14},
David~N.~Spergel\altaffilmark{2},
Suzanne~T.~Staggs\altaffilmark{11},
Eric~R.~Switzer\altaffilmark{15,4},
Hy~Trac\altaffilmark{3},
Edward~J.~Wollack\altaffilmark{15}
}

\altaffiltext{1}{Sub-department of Astrophysics, University of Oxford, Keble Road, Oxford OX1 3RH, UK}
\altaffiltext{2}{Department of Astrophysical Science, Peyton Hall, 4 Ivy Lane, Princeton NJ 08544, USA}
\altaffiltext{3}{McWilliams Center for Cosmology, Carnegie Mellon University, Pittsburgh PA 15213, USA}
\altaffiltext{4}{Canadian Institute for Theoretical Astrophysics, University of Toronto, Toronto, ON, Canada M5S 3H8}
\altaffiltext{5}{Department of Physics, Cornell University, Ithaca, NY 14853, USA}
\altaffiltext{6}{Department of Physics and Astronomy, University of Pennsylvania, 209 South 33rd Street, Philadelphia, PA, USA 19104}
\altaffiltext{7}{Department of Physics and Astronomy, University of Pittsburgh, Pittsburgh, PA, USA 15260}
\altaffiltext{8}{Department of Physics, University of Michigan Ann Arbor, MI 48109, USA}
\altaffiltext{9}{Astrophysics and Cosmology Research Unit, School of Mathematics, Statistics and Computer Science, University of KwaZulu-Natal, Durban 4041, South Africa}
\altaffiltext{10}{Dunlap Institute for Astronomy and Astrophysics, University of Toronto, 50 St. George St., Toronto, ON, M5S 3H4, Canada}
\altaffiltext{11}{Joseph Henry Laboratories of Physics, Jadwin Hall, Princeton University, Princeton, NJ 08544, USA}  
\altaffiltext{12}{Department of Physics and Astronomy, Haverford College, Haverford, PA 19041, USA}
\altaffiltext{13}{Physics and Astronomy Department, Stony Brook University, Stony Brook, NY USA 11794}
\altaffiltext{14}{Astrophysics and Cosmology Research Unit, School of Chemistry and Physics, University of KwaZulu-Natal, Durban 4041, South Africa}
\altaffiltext{15}{NASA/Goddard Space Flight Center, Greenbelt, MD, USA 20771}
 
\begin{abstract}

Future arcminute resolution polarization data from ground-based Cosmic Microwave Background (CMB) observations can be used to estimate the contribution to the temperature power spectrum from the primary anisotropies and to uncover the signature of reionization near $\ell=1500$ in the small angular-scale temperature measurements. Our projections are based on combining expected small-scale E-mode polarization measurements from Advanced ACTPol in the range $300<\ell<3000$ with simulated temperature data from the full Planck mission in the low and intermediate $\ell$ region, $2<\ell<2000$.
We show that the six basic cosmological parameters determined from this combination of data will predict the underlying primordial temperature spectrum at high multipoles to better than $1\%$ accuracy. Assuming an efficient cleaning from multi-frequency channels of most foregrounds in the temperature data, we investigate the sensitivity to the only residual secondary component, the kinematic Sunyaev-Zel'dovich (kSZ) term. The CMB polarization is used to break degeneracies between primordial and secondary terms present in temperature and, in effect, to remove from the temperature data all but the residual kSZ term. We estimate a $15 \sigma$ detection of the diffuse homogeneous kSZ signal from expected AdvACT temperature data at $\ell>1500$, leading to a measurement of the amplitude of matter density fluctuations, $\sigma_8$, at $1\%$ precision. Alternatively, by exploring the reionization signal encoded in the patchy kSZ measurements, we bound the time and duration of the reionization with $\sigma(\zre)=1.1$ and $\sigma(\delz)=0.2$. We find that these constraints degrade rapidly with large beam sizes, which highlights the importance of arcminute-scale resolution for future CMB surveys.

\end{abstract}


\section{Introduction}
\label{sec:intro}

The Cosmic Microwave Background (CMB) anisotropies have proven to be some of the most powerful probes of the early universe. 
Most recently, the standard cosmological model has been constrained with high precision by data from the Planck mission \citep{pi,pxv,pxvi}, the 9-year data release from the Wilkinson Microwave Anisotropy Probe \citep[WMAP,][]{Bennett:2012zja,Hinshaw:2012aka}, and high-resolution measurements from the Atacama Cosmology Telescope \citep[ACT,][]{das2013,dunkley2013,sievers2013,calabrese2013} and the South Pole Telescope \citep[SPT,][]{Story:2012wx,Hou:2012xq}.
While the CMB temperature anisotropies, both the primordial contributions and the late-time secondary components, have been measured with high precision over a wide range of angular scales, we are entering a new and challenging phase devoted to characterizing the CMB polarization signal. To date, the even-parity E-mode polarization signal has been fixed by WMAP low-$\ell$ data \citep{Bennett:2012zja} followed at degree scales by DASI \citep{dasi}, BOOMERanG \citep{boomerang, boomerang2}, CBI \citep{cbi}, CAPMAP \citep{capmap}, QUAD \citep{quad}, QUIET \citep{quiet}, BICEP1 \citep{bicep1}, BICEP2 \citep{bicep2}, POLARBEAR \citep{PBpol} and ACTPol \citep{actpol} measurements. More precise characterization of the large and intermediate scale ($2<\ell<2500$) E-modes is expected soon from the Planck team. Evidence for the B-modes, the most challenging observable to measure, was presented recently by BICEP2 \citep{bicep2} at low multipoles, characterizing primordial polarization, and by the SPTPol and POLARBEAR collaborations \citep{hanson,Ade:2013gez,Ade:2013hjl,PBpol} at intermediate-to-high multipoles, who presented a measurement of the B-modes produced by gravitational lensing.

The next decade will see further improvements in ground-based instruments designed to measure the polarization of the microwave sky with high precision. These observations would be complemented by a future space mission \citep[e.g.,][]{epic, core, pixie, litebird, prism} expected to supply cosmic variance limited measurements of the polarization spectrum at very large scales. As recently highlighted in \citet{Galli:2014kla}, CMB polarization will provide an important consistency test for temperature observations and will help to pin down cosmological parameters. 

In this work we estimate the constraining power of arcminute resolution polarization data and make use of it to break degeneracies between primordial and secondary terms present in temperature anisotropies. We investigate how, by estimating the primordial cosmology with expected small-scale polarization data, one can for the first time use the measured temperature fluctuations at $\ell>1500$ to directly probe the wide range of astrophysical processes affecting the temperature spectrum after the epoch of decoupling. In particular, we focus on the only source of secondary temperature anisotropies degenerate in frequency space with the CMB, namely the kinematic Sunyaev-Zel'dovich (kSZ) effect \citep{Suny1980}. This method can be more generally applied to any non-parametric component which does not correlate between temperature and polarization.

We anticipate many small-scale measurements of CMB polarization in the coming years. As an example, we discuss Advanced ACTPol (AdvACT), a new receiver for the Atacama Cosmology Telescope designed to probe small-scale polarization.
ACT is a 6-meter off-axis Gregorian telescope located on Cerro Toco in the Atacama desert of northern Chile, at an elevation of 5190 meters. The ACT temperature survey operated from 2008 until 2010 \citep{swetz/etal/2011}. The instrument was subsequently upgraded and is currently observing with the ACTPol receiver \citep{Niemack:2010wz}. Based on projected improvements of the existing ACTPol camera we consider a possible extension of the ACTPol project, AdvACT, providing a combination of high resolution and sensitivity, and wide frequency and sky coverage. AdvACT is designed to address a wide range of cosmological goals from primordial B-modes to SZ science. A complete description will appear in a future paper, in this work we concentrate on the science coming from enabling a cosmic variance limited measurement of the E-mode polarization spectrum out to $\ell \approx 2000$.

The paper is organized as follows. In Sec.~\ref{sec:physics} we review the basic physics of CMB polarization and its coupling with CMB temperature. In Sec.~\ref{sec:data} we describe simulations of temperature and polarization observations based on predictions of the full-mission Planck sensitivity and the expected AdvACT EE sensitivity. Sec.~\ref{sec:predicting} reports the derived prediction of the high-$\ell$ temperature power spectrum. In Sec.~\ref{sec:ksz} we investigate the sensitivity of the temperature data at high multipoles to constrain secondary anisotropies, focusing on the kinematic Sunyaev-Zel'dovich component, and using it to probe the reionization history of the universe. We discuss the results and conclude in Sec.~\ref{sec:disc}.

\section{CMB Temperature-Polarization Coupling}
\label{sec:physics}
The balance between gravity and photon pressure generates acoustic oscillations in the photon-baryon plasma in the early universe.  According to the gravitational instability paradigm, the CMB temperature anisotropies caused by the acoustic oscillations are sourced by primordial fluctuations in the density of the early universe. The anisotropies we observe at the recombination epoch then encode integrated information about the density perturbations that are produced at reheating and grow until decoupling. In addition, the variations in the plasma velocities at decoupling leads, via Thomson scattering, to a pattern of linear polarization on the CMB anisotropies themselves \citep[e.g.,][]{Bond1984,Zaldarriaga:1995gi,Kamionkowski:1996ks,seljak96,Zaldarriaga:1996xe,Hu:1997hp,Hu:1997hv}. The degree of linear polarization is proportional to the local quadrupole anisotropy of the incident photons when they last scatter \citep{Zaldarriaga:1995gi,Hu:1997hp,Hu:1997hv}. 

As long as the CMB photons are tightly coupled to the charged electrons, only monopole and dipole fluctuations are generated. A quadrupole moment arises later through free-streaming as photons begin to decouple very near the surface of last scattering. CMB polarization therefore probes a narrow redshift window, directly characterizing the time and duration of the decoupling epoch. 
The evolution of the free electron density during late-time cosmic reionization also imprints an anisotropy in the CMB polarization pattern \citep[see e.g.][]{Hu:1997hv}.

In contrast to observations of temperature anisotropies, CMB polarization measurements have the power to isolate and characterize specific epochs, such as the epochs of recombination and late-time reionization, because the polarization depends on the free-electron density at that epoch. Measurements of CMB polarization will then help in constraining cosmological parameters affecting those epochs and hence break degeneracies present when considering only temperature data, increasing the precision of cosmological estimates \citep[see][and references therein]{Galli:2014kla}. 
Moreover, the exact properties of the polarization pattern depend on the mechanism that produced the temperature anisotropies, which aids in distinguishing the scalar, vector, or tensor nature of the primordial fluctuations.

In the presence of scalar perturbations, only temperature anisotropies and E-modes of polarization are produced. A non-zero coupling then holds only between the T and E field, while the curl-type B-mode component vanishes.
The temperature and E polarization spectra are given by (see \citet{Zaldarriaga:1996xe} and references therein)
\begin{eqnarray}
C^{TT}_\ell &=& (4\pi)^2\int k^2dkP(k)\left[\Delta^{TT}_\ell(k,\eta_0)\right]^2 \nonumber \\
C^{EE}_\ell &= &(4\pi)^2\int k^2dkP(k)\left[\Delta^{EE}_\ell(k,\eta_0)\right]^2 \,,
\end{eqnarray}
where $P(k)$ is the primordial three-dimensional power spectrum in terms of wavenumber $k$. $\Delta^{TT}_\ell(k,\eta_0)$ and $\Delta^{EE}_\ell(k,\eta_0)$ are the transfer functions at the present conformal time $\eta_0$, written as
\begin{eqnarray}
\Delta^{TT}_\ell(k,\eta_0)  &=& \int_0^{\eta_0}d\eta S_{T}(k,\eta)j_\ell(k(\eta_0-\eta))\nonumber\\
\Delta^{EE}_\ell(k,\eta_0)  &=&  \sqrt{\frac{(\ell+2)!}{(\ell-2)!}}\int_0^{\eta_0}d\eta S_{E}(k,\eta)j_\ell(k(\eta_0-\eta))\,,\nonumber\\
\label{eq:transfer}
\end{eqnarray}
with geometric dependence encoded by the $j_\ell$ spherical Bessel functions. Cosmological dependence is encoded by $S_T(k,\eta), S_E(k, \eta)$, the source terms of the temperature and E-mode polarization respectively.

The temperature and polarization transfer functions are themselves strong functions of the underlying cosmological parameters, and hence each independently constrains the underlying cosmology.
\begin{figure}[t!]
\includegraphics[width=\columnwidth]{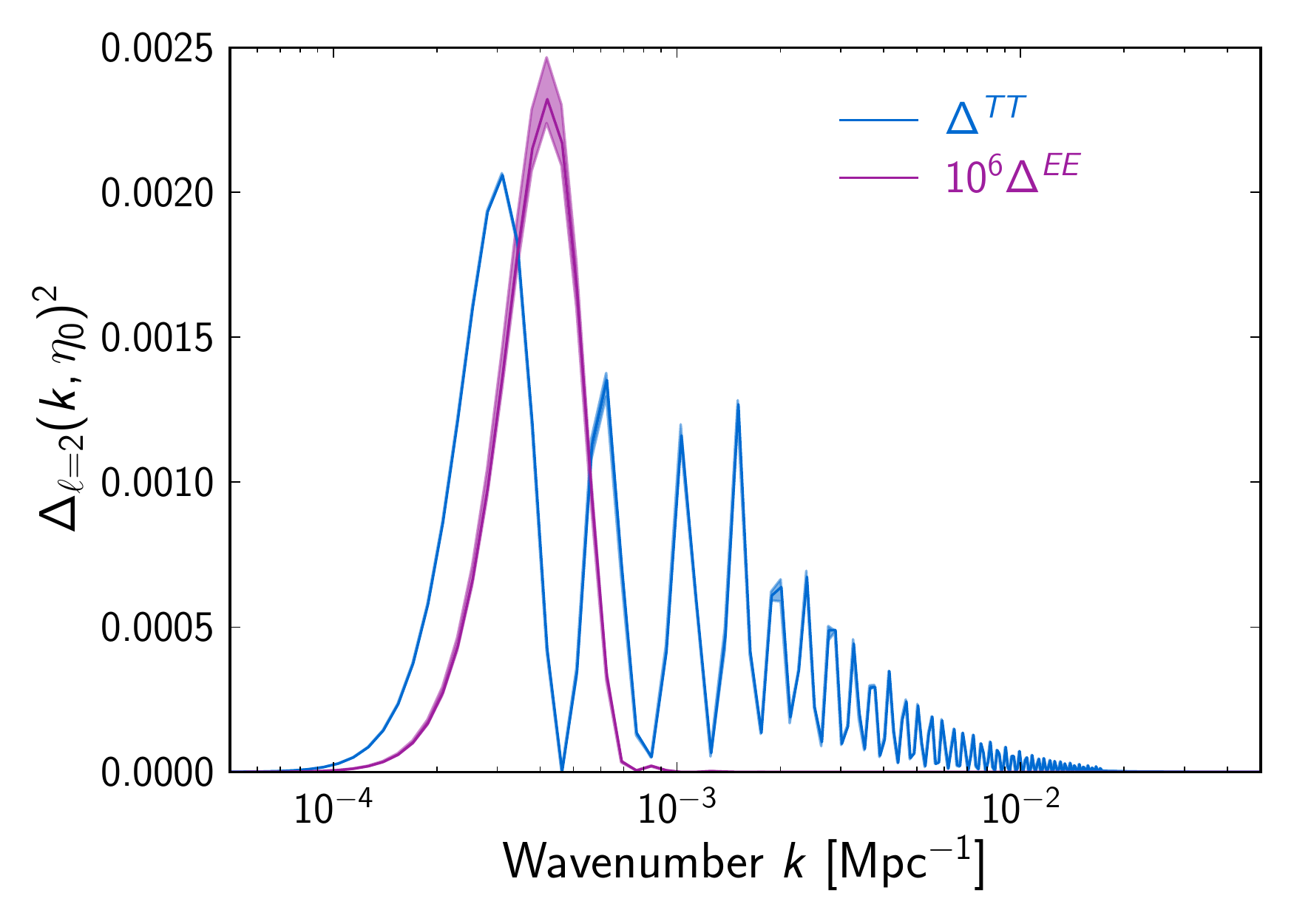}
\caption{Temperature (blue) and polarization (purple) quadrupole transfer functions. The solid lines show the transfer functions corresponding to the Planck 2013 {\em base\_planck\_lowl\_lowLike\_highL} best-fit cosmological model, while the bands show the transfer functions computed for $\Omega_bh^2$ at the left- and right-most $2\sigma$ range allowed by the Planck data. This shows how particularly the polarization transfer function constrains the cosmological model, allowing one to predict the temperature transfer function (and thus the angular power spectrum) using EE data.\\
\label{fig:transfer}
}
\end{figure}

Each source term in Eq.~\ref{eq:transfer} is the same for all multipoles and only depends on photon moments up to $\ell=4$ \citep{seljak96}.
Specifically, the derivation in \citet{Zaldarriaga:1995gi} shows that in the tight coupling approximation $S_E(k, \eta)$ can be directly related to the temperature quadrupole which is, in turn, produced by the free-streaming of the temperature dipole in the last few photon-electron scatterings, and is proportional to the temperature dipole and the width of the last scattering surface. By constraining the polarization transfer functions one then can in fact `reverse' this process and use the CMB  E-mode polarization pattern as a tracer of the underlying primordial temperature anisotropies. That is, the cosmology determined from the polarization power spectrum in effect predicts the temperature power spectrum.
 Fig.~\ref{fig:transfer} shows the TT and EE quadrupole transfer functions, both fully specified by a given cosmological model.

An advantage of this method is that, although several galactic and extragalactic foregrounds contribute significantly to temperature anisotropies, a small fraction of sources is polarized leading to polarization measurements, and a cosmological reconstruction, that are less contaminated.

\section{Simulated Data}
\label{sec:data}
\begin{figure}[t!]
\center
\includegraphics[width=\columnwidth]{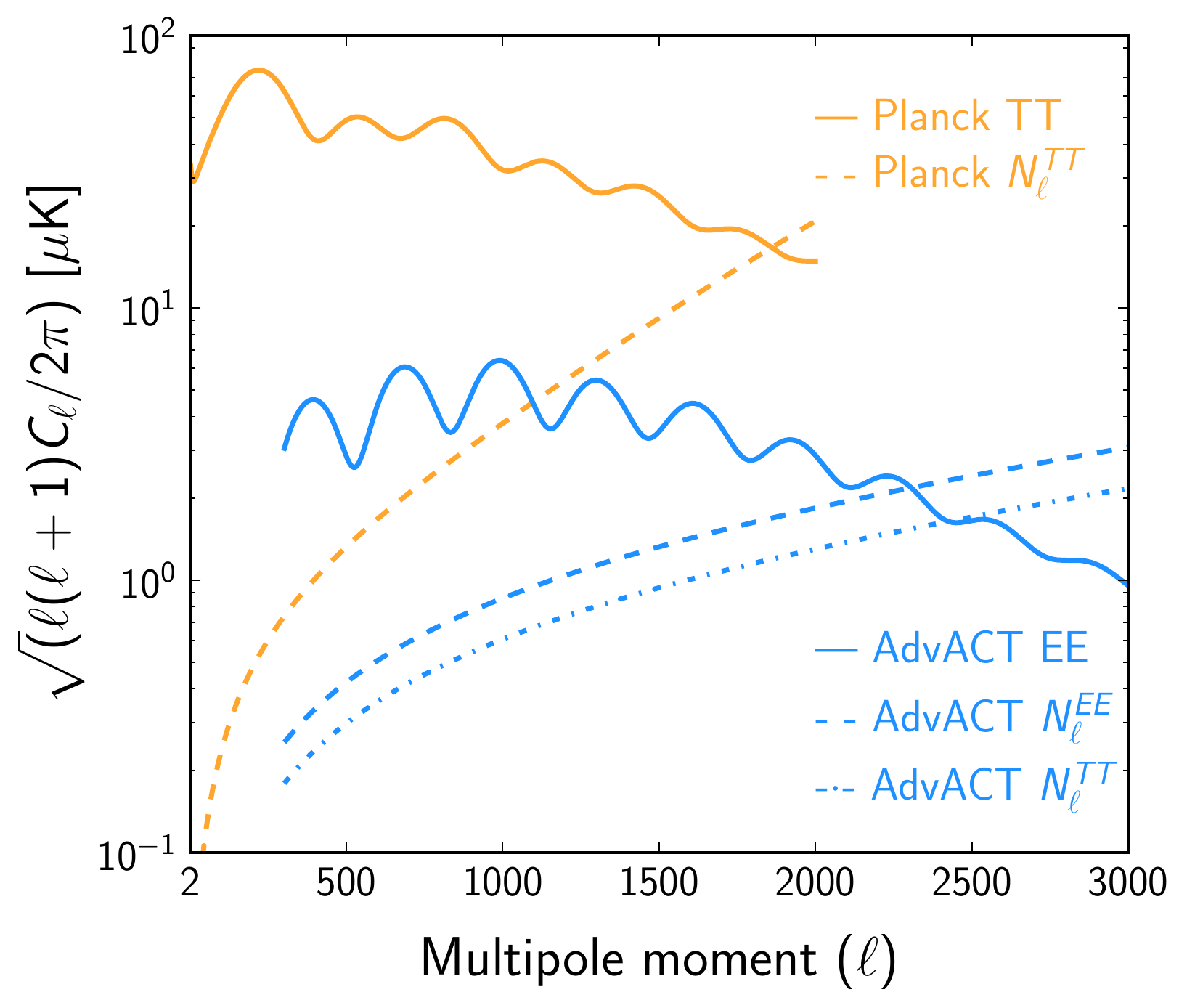}
\caption{Simulated Planck temperature and AdvACT polarization observations with relative noise levels.\\
\label{fig:data}
}
\end{figure}

\begin{table}[t!]
\begin{center}
\begin{tabular}{lcccc}
\hline
\hline
Experiment & Channel & $\Theta_{\rm FWHM}$ & Map Noise\\
& [GHz] & [arcmin] & [$\mu$K$-{\rm arcmin}$] \\
\hline
& & & & \\
Planck & 70 & 14 & 180 \tablenotemark{a}\\
\phantom{Planck} & 100 & 10 & 65\\
\phantom{Planck} & 143 & 7.1& 43\\
\phantom{Planck} & 217 & 5.0& 65\\
f$_{\rm sky}$=0.8 & & & \\
\hline
&&&& \\
AdvACT & 90 & 2.2 & 7.8 \\
\phantom{Planck} & 150 & 1.3 & 6.9\\
\phantom{Planck} & 230 & 0.9 & 25 \\
f$_{\rm sky}$=0.5 & & & \\
\hline
\hline
\footnotetext[1]{Planck nominal mission map noise. We use scaled versions of Planck's pre-launch estimates of sensitivity, well aware that the full mission, in flight, values may differ from them to some degree.} 
\end{tabular}
\caption{Projected Planck and AdvACT properties used for the estimates in this paper.}
\label{table:exp}
\end{center}
\end{table}
It is anticipated that AdvACT will report a cosmic variance-limited measurement of the E-mode polarization spectrum out to $\ell \approx 2000$ observing half the sky (see Fig.~\ref{fig:data}). In this work we consider AdvACT EE measurements in the range $300<\ell<3000$ and supplement this data with the expected full mission Planck temperature measurements in the range $2<\ell<2000$. We do not include Planck polarization data as this work focuses on the power of arcminute-resolution EE data to predict the small-scale temperature spectrum.

We simulate the Planck full mission TT and AdvACT EE polarization spectra, using a reference fiducial cosmology based on the {\em base\_planck\_lowl\_lowLike\_highL} from the Planck 2013 results \citep{pxvi}. We rescale the Planck nominal sensitivities \citep{blueplanck} to the expected full mission performances considering $\sim$30 months of observation \citep{pi}; we consider only the combination of frequency channels optimal for cosmological analyses: $70, 100, 143$ and $217$ GHz, and we adopt a sky coverage of 80\%. For the AdvACT E-mode projected data we assume a total scan area of 20000 deg$^2$ (corresponding to a sky fraction f$_{\rm sky}=0.5$) across the cosmological frequency channels $90, 150$ and $230$ GHz. To evaluate the AdvACT polarization sensitivities we rescale the temperature sensitivities by a factor of $\sqrt{2}$. The experimental specifications for both experiments are reported in Table~\ref{table:exp}, while Fig.~\ref{fig:data} shows the spectra and the noise levels for the simulated data. For a single frequency, the error for the $\ell^\mathrm{th}$ bin is
\begin{equation}
N_{\ell}=\sigma_0^2\exp(\ell(\ell+1)\Theta^2_{\rm FWHM}/(8 \ln2))/(4\pi) \, 
\end{equation} 
where $\sigma_0$ is the map noise and $\Theta_{\rm FWHM}$ is the beam full width half-maximum.

While there is a non-negligible overlap in sky coverage between the two experiments, we estimate the correlations between the AdvACT EE and Planck TT spectra to be less than $10\%$ and neglect them in this analysis. Moreover, ignoring correlations between TT and EE is in fact more conservative as those correlations would reduce the cosmic variance and may lead to tighter constraints.

In Sec.~\ref{sec:ksz} we will introduce simulated AdvACT TT observations, selecting $300<\ell<3000$. The corresponding temperature noise for AdvACT is then also shown in Fig.~\ref{fig:data}.

\section{Predicting Temperature from Polarization}
\label{sec:predicting}

Assuming that multi-frequency observations will provide a foreground cleaned temperature dataset and that the small-scale polarization data will contain negligible contamination from foregrounds, in this Section we constrain the contribution to the TT power spectrum from primordial anisotropies without accounting for residual foreground levels. That is, we predict a foreground-free temperature spectrum from a measured polarization spectrum. Our assumption is further supported by the selected $\ell$ range in both temperature and polarization data. The primary CMB temperature fluctuations dominate the microwave measurements at $\ell<2000$ \citep{reichardt2011,dunkley2013,pxvi}. In polarization we instead expect polarized galactic dust contamination at very low multipoles ($\ell<300$) and emission from polarized Poisson sources at very high multipoles ($\ell>3000$) \citep{actpol}.

We consider a spatially flat $\Lambda$CDM base model described by six cosmological parameters: the baryon and cold dark matter densities, $\Omega_b h^2$ and $\Omega_c h^2$; the angular scale of the acoustic horizon at decoupling, $\theta_{\rm A}$; the reionization optical depth, $\tau$; and the amplitude and the scalar spectral index of primordial adiabatic density perturbations, $A_{\rm s}$  and $n_{\rm{s}}$ at a pivot scale $k_0 = 0.05~\rm{Mpc}^{-1}$. We use the dataset described above to estimate the aforementioned cosmological parameters, using the standard MCMC technique and calling the exact likelihood routine implemented in the publicly available \texttt{cosmomc} \citep{Lewis:2002ah} package. 

\begin{figure}[t!]
\includegraphics[width=\columnwidth]{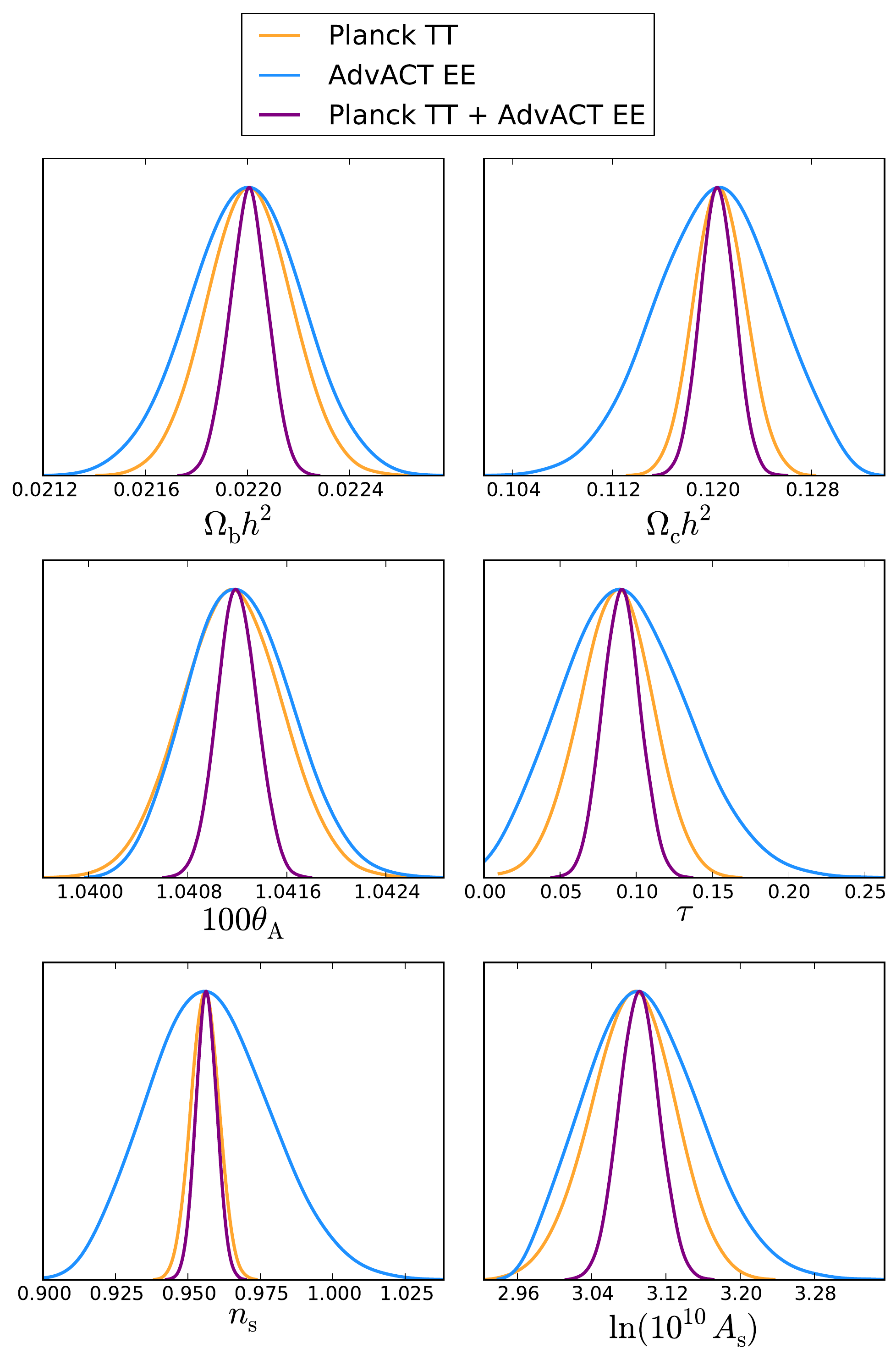}
\caption{Posterior distributions of the base $\Lambda$CDM cosmological parameters derived from simulated Planck temperature data alone (orange), AdvACT EE data (blue), and their combination (purple).\\
\label{fig:cosmo}
}
\end{figure}

In Fig.~\ref{fig:cosmo} we compare the estimated posterior distributions of the base cosmological parameters. Note that since we do not include low-$\ell$ polarization data, the EE data alone do not constrain $n_{\rm{s}}$ and $\Omega_c h^2$ well. However, the clean measurement of the high-order acoustic peaks in EE traces out the overall shape of the spectrum well and is thus especially good in constraining $\theta_{\rm A}$, fixed by the positions of the peaks. Notably,
the accurate measurement of the high-$\ell$ damping region of the spectrum, and hence of the CMB gravitational lensing effect, helps in breaking the degeneracy between $A_s$ and $\tau$, for both Planck and AdvACT. More generally, considering both datasets leads to a factor of roughly $\sqrt2$ improvement in parameter constraints, because most parameters are dominated by cosmic variance at these sensitivities. Similar conclusions were drawn in \citet{rocha} and \citet{Galli:2014kla}. 

\begin{figure*}[htb!]
\begin{flushright}
\begin{tabular}{cc}
\includegraphics[scale=0.5]{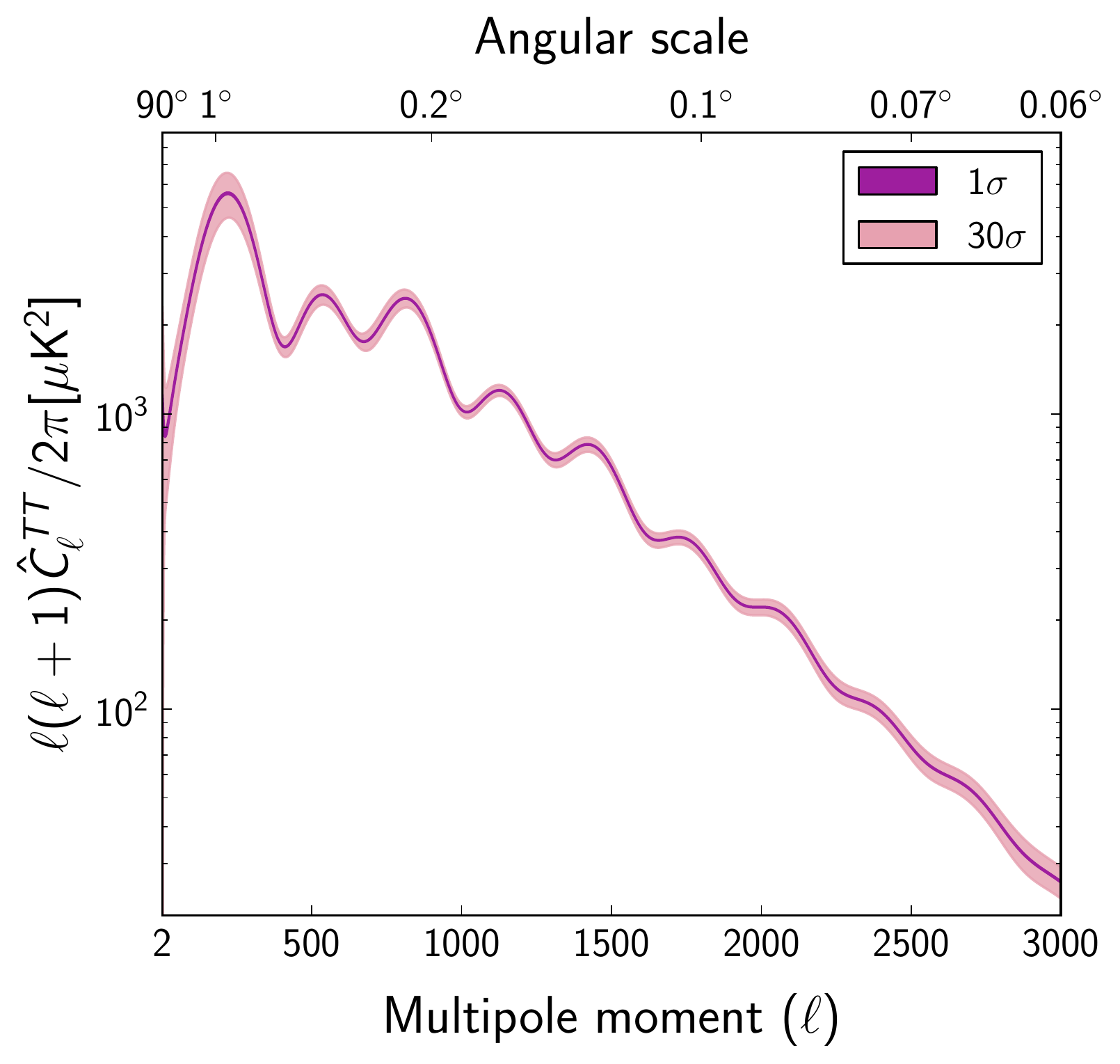}&
\includegraphics[width=\columnwidth]{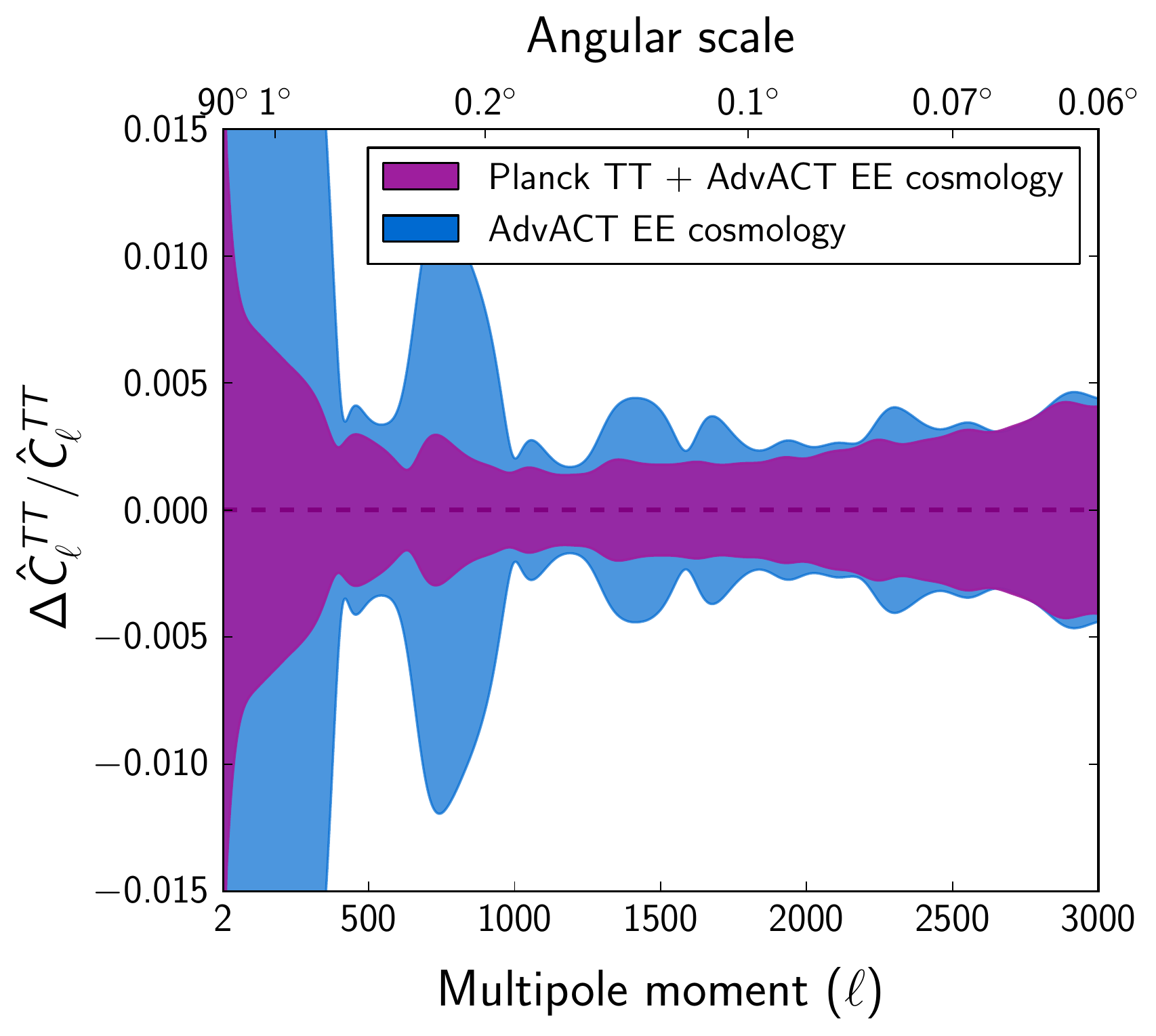}
\end{tabular}
\caption{\emph{Left}: Predicted contribution to the temperature power spectrum from primordial anisotropies obtained from the Planck TT + AdvACT EE cosmology. We report the 1$\sigma$ band uncertainty of the mean extracted temperature and, for a better visualization, the same band enlarging the errors by a factor of 30. \emph{Right}: zoom on the 1$\sigma$ band uncertainty after subtracting the mean extracted $\hat{C}^{TT}_{\ell}$ per $\hat{C}^{TT}_{\ell}$, i.e. the fractional error in the recovered  $\hat{C}^{TT}_{\ell}$. We compare the Planck TT + AdvACT EE predicted error on $\hat{C}^{TT}_{\ell}$ with the same error band obtained using AdvACT EE only cosmology. At small scales, $\ell>2200$, the estimate is dominated by AdvACT.
\label{fig:predictedtt}
}
\end{flushright}
\end{figure*}
\subsection{Predicting the underlying primary TT power}

In addition to simply constraining cosmological parameters we take a further step and use the inferred cosmology to predict the underlying primary temperature power that a small-scale polarization experiment should see. We generate a predicted temperature spectrum, $\hat{C}^{TT}_\ell$, by estimating the primordial temperature spectrum for each $\ell$ at each step in the MCMC Planck TT + AdvACT EE chain. We then extract the mean and covariance of the $\hat{C}^{TT}_\ell$ chain, as shown in Fig.~\ref{fig:predictedtt}. The mean $\hat{C}^{TT}_{\ell}$ is shown with a 1$\sigma$ band of uncertainty (the diagonal of the covariance) in the left panel; the constraint is so tight that it can only be seen by enlarging the error band, as shown with the 30$\sigma$ band. The covariance between the  $\hat{C}^{TT}_\ell$ at adjacent $\ell$ values vanishes, given that we are using a large fraction of the sky. In the right panel we zoom on the 1$\sigma$ band uncertainty by presenting the mean-subtracted band per $\hat{C}^{TT}_{\ell}$. The predicted underlying CMB temperature is bounded within a $0.5\%$ uncertainty region at high multipoles, where AdvACT dominates the constraints.

In the right hand panel of Fig.~\ref{fig:predictedtt} we compare the 1$\sigma$ uncertainty band described above with the equivalent band obtained using a cosmological model derived from the AdvACT EE measurements only (e.g., ignoring Planck TT data). As expected the $\hat{C}^{TT}_{\ell}$ spectrum is poorly constrained at low multipoles ($\ell<300$), due to the fact that we do not consider EE data on large scales\footnote{AdvACT is expected to report data below $\ell = 300$ but in this paper we considered only small-scale measurements.}. The constraining power of AdvACT increases with multipole, at high multipoles, $\ell > 2200$, the spectrum uncertainty is dominated by the EE polarization constraints.

\subsection{Tests of robustness}
In order to confirm the robustness of our technique, we consider different combinations of data and theoretical assumptions made in performing the $\hat{C}^{TT}_\ell$ prediction.
\begin{itemize}
\item As mentioned above we tested for the dependence of the extraction on the Planck dataset. We have shown in the right panel of Fig.~\ref{fig:predictedtt} that the TT small-scale predictions are dominated by AdvACT EE data at multipoles $\ell > 2200$.

\item We checked for the dependence of the extraction on gravitational lensing of the CMB by analysing only unlensed scalar power spectra. Assuming a perfect de-lensing pre-processing of the data, one would expect a more definite measurement of the amplitude and position of the acoustic peaks. However, when assuming unlensed spectra in our extraction of $\hat{C}^{TT}_{\ell}$, we find a negligible improvement in the extracted spectrum over the whole angular range. The error on the predicted power spectrum increases at most by $0.2\sigma$ when including the effects of lensing. Switching the lensing smearing of the acoustic peaks on or off will increase or decrease the error band slightly but will not bias the reconstructed spectrum.
 
\item We checked for the dependence of the extracted spectrum on the cosmological model. 
\begin{itemize}
\item In addition to the base $\Lambda$CDM model, we considered a one parameter extension by letting the effective number of relativistic species, $N_{\rm eff}$, vary. The relativistic particle content affects the CMB high-$\ell$ damping tail and hence will increase the uncertainty in the AdvACT-dominated region. In this case, the 1$\sigma$ estimated uncertainty band broadens by about 10-30\% for multipoles in the range $1000<\ell< 3000$ (see the blue band in Fig.~\ref{fig:extband}).
\item In light of the recent results from BICEP2 \citep{bicep2} we also opened the parameter space allowing the tensor-to-scalar ratio $r$ and the running of the spectral index to vary. As expected (see the orange band in Fig.~\ref{fig:extband}), adding these two parameters increases the uncertainty at very low multipoles ($\ell<200$), with no effect on the science case at high-$\ell$ as investigated in this paper.
\end{itemize}

\begin{figure}[t!]
\includegraphics[width=\columnwidth]{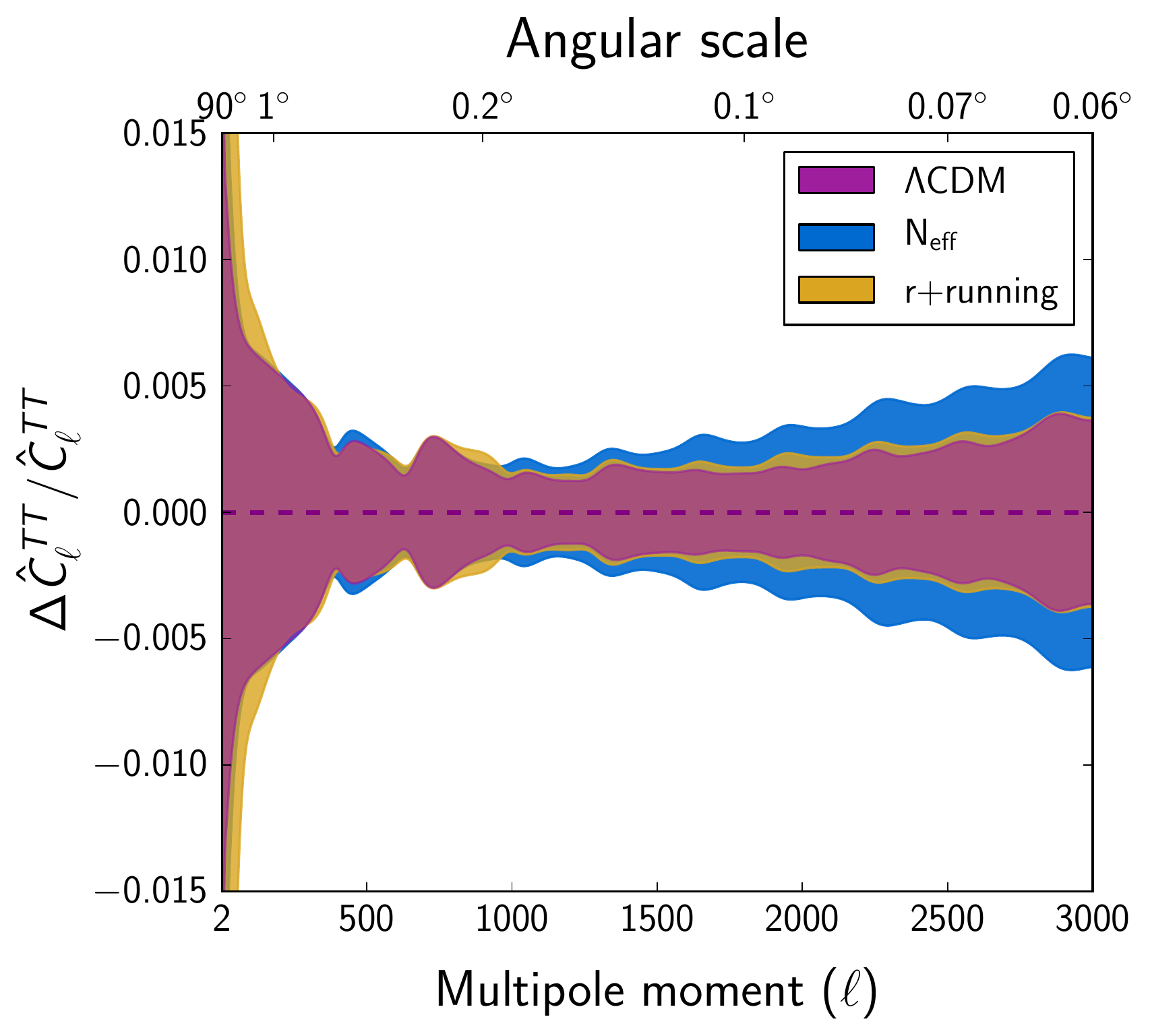}
\caption{Fractional error band on the predicted $\hat{C}^{TT}_{\ell}$ from the Planck TT + AdvACT EE cosmology comparing the predicted $\hat{C}^{TT}_{\ell}$ assuming the standard $\Lambda$CDM model (purple band) or: varying the effective number of relativistic species N$_{\rm eff}$ (blue band); varying the tensor-to-scalar ratio $r$ and the running of the spectral index (orange band).\\
\label{fig:extband}
}
\end{figure}

\end{itemize}

\section{Revealing the Kinematic Sunyaev-Zel'dovich Spectrum}
\label{sec:ksz}
Estimating the contribution to the temperature power spectrum from primordial anisotropies that a small-scale CMB experiment would infer from its polarization data allows one to break degeneracies between primordial and secondary contributions present in temperature observations, and to directly probe the wide range of astrophysical processes affecting the temperature spectrum at high multipoles.
Broad frequency coverage is a feature of both the Planck and expected AdvACT results, and is crucial to reduce the contamination from astrophysical foregrounds, such as the cosmic infrared background \citep[CIB, e.g.,][]{Puget1996}, radio point sources, the thermal Sunyaev-Zel'dovich (tSZ) effect \citep{Suny1970}, and galactic dust and synchrotron emission. All these foregrounds can be separated out from the primordial CMB anisotropies because they have spectral signatures that differ from the CMB. The only residual component completely degenerate in frequency space with the CMB is the diffuse kinematic Sunyaev-Zel'dovich (kSZ) emission \citep{Suny1980}, which has a blackbody spectrum with a predicted contribution at $\ell=3000$ of a few $\mu {\rm K}^2$ \citep{reichardt2011,dunkley2013,sievers2013,pxvi}.

The SZ effect is caused by the inverse-Compton scattering of low energy CMB photons off the high energy electrons in galaxy clusters. While the tSZ effect results from scattering off hot electrons and depends on integrated electron pressure along the line of sight, the
kSZ effect is due to the peculiar motion of the electrons and scales with the electron momentum.
The kSZ power is expected to have a low-redshift contribution, namely the homogeneous kSZ, sourced by perturbations in the free electron density after reionization and caused by the peculiar velocity of the intergalactic medium and unresolved galaxy clusters \citep{ostriker/vishniac:1986}. An additional high-redshift component is sourced by fluctuations in the ionized fraction as well as the electron density during reionization. We call this high-redshift component the patchy kSZ \citep[e.g.,][]{Knox:1998fp,gruzinov/hu:1998,mcquinn/etal:2005,iliev/etal:2007}. 

\subsection{kSZ extraction}
We now want to extend our temperature dataset to investigate secondary anisotropies affecting the TT spectrum at high multipoles. Using the specifications introduced in Sec.~\ref{sec:predicting} we then take as temperature measurements the expected small-scale AdvACT TT data, in the range $300<\ell<3000$. As we are using higher multipoles, we also consider foreground contamination in the TT spectrum and include emission from the kSZ effect. As mentioned before, assuming efficient cleaning from observations at many frequencies, the kSZ is the only residual secondary component present in the data. 
To reveal the kSZ power present in small-scale temperature measurements we: i) consider the kSZ to give a negligible contribution to the polarization and assume that the EE spectrum predicts the primary high-$\ell$ TT as shown in the previous Section; ii) constrain at the same time the primary CMB fluctuations and the kSZ contributions using simulated AdvACT temperature and polarization observations.
The constraining power for primordial CMB coming from EE will break degeneracies between primordial and secondary temperature anisotropies and then unveil the kSZ terms. The kSZ extraction is possible given that the remaining underlying primary CMB uncertainty is negligible in power compared to kSZ at $\ell>1500$. 
\begin{figure}[t!]
\includegraphics[width=\columnwidth]{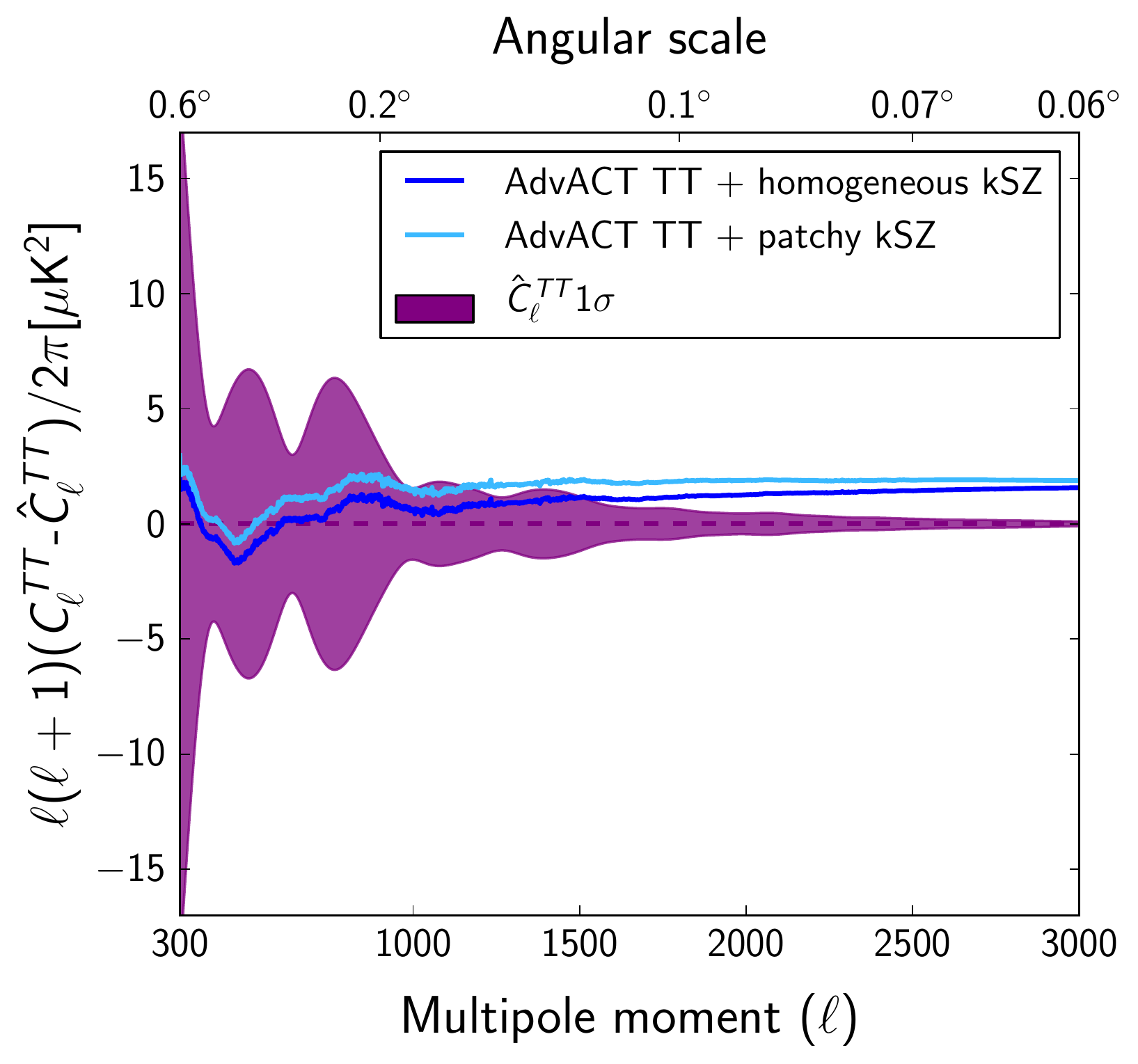}
\caption{Residual power in the simulated AdvACT TT + kSZ components with respect to the predicted TT from the Planck TT + AdvACT EE cosmology. The blue lines are one realization of the data including homogeneous (dark blue) or patchy (light blue) terms, while the purple band is the error on the predicted TT spectrum derived in the previous Section. At multipoles $\ell>1500$ the kSZ signal is well above the error on the predicted primordial TT spectrum.\\
\label{fig:kszres}
}
\end{figure}
We show this in Fig.~\ref{fig:kszres}, where we report the residual power between one realization of the AdvACT TT spectrum and the kSZ signal (allowing for two models) with respect to the predicted primary CMB-only TT power extracted from the Planck TT + AdvACT EE cosmology. The kSZ clearly dominates over the CMB-only at high multipoles. 

We include two contributions for the kSZ, namely:
\begin{enumerate}
\item \textit{Homogeneous kSZ}\\
This first low-redshift component assumes a model with instantaneous reionization, presented in \citet{Battaglia:2010tm} and derived from hydrodynamic simulations described in \citet{Battaglia:2011cq}.  The predicted amplitude of the signal in $\ell(\ell+1)C_\ell/2\pi$ at $\ell=3000$ from the simulations is h-A$_{\rm kSZ}=1.5\ \mu$K$^2$ for homogeneous reionization at $z=10$ in a $\sigma_8=0.8$ cosmology, the value of $\sigma_8$ assumed in our simulations. 
We note that the amplitude of this template is sensitive to the characteristics and assumptions of the simulations. For example, if we follow the scaling relations derived in \citet{Shaw:2011sy}, which account for simulation box size and for a change in the instantaneous redshift of reionization (from $z=10$ down to $z=5.5$), then we find h-A$_{\rm kSZ}$ = 1.65 $\mu$K$^2$ for this particular homogeneous kSZ template.

\item \textit{Patchy kSZ}\\
We also consider the patchy kSZ signal at high redshift described in \citet{Battaglia:2012im} where the assumption of instantaneous reionization is relaxed. The patchy template has a predicted amplitude at $\ell=3000$ of p-A$_{\rm kSZ}=1.8\ \mu$K$^2$, which accounts for all the $z>5.5$ kSZ contributions, and no additional scaling is needed.
\end{enumerate}

\begin{figure}[t!]
\includegraphics[width=\columnwidth]{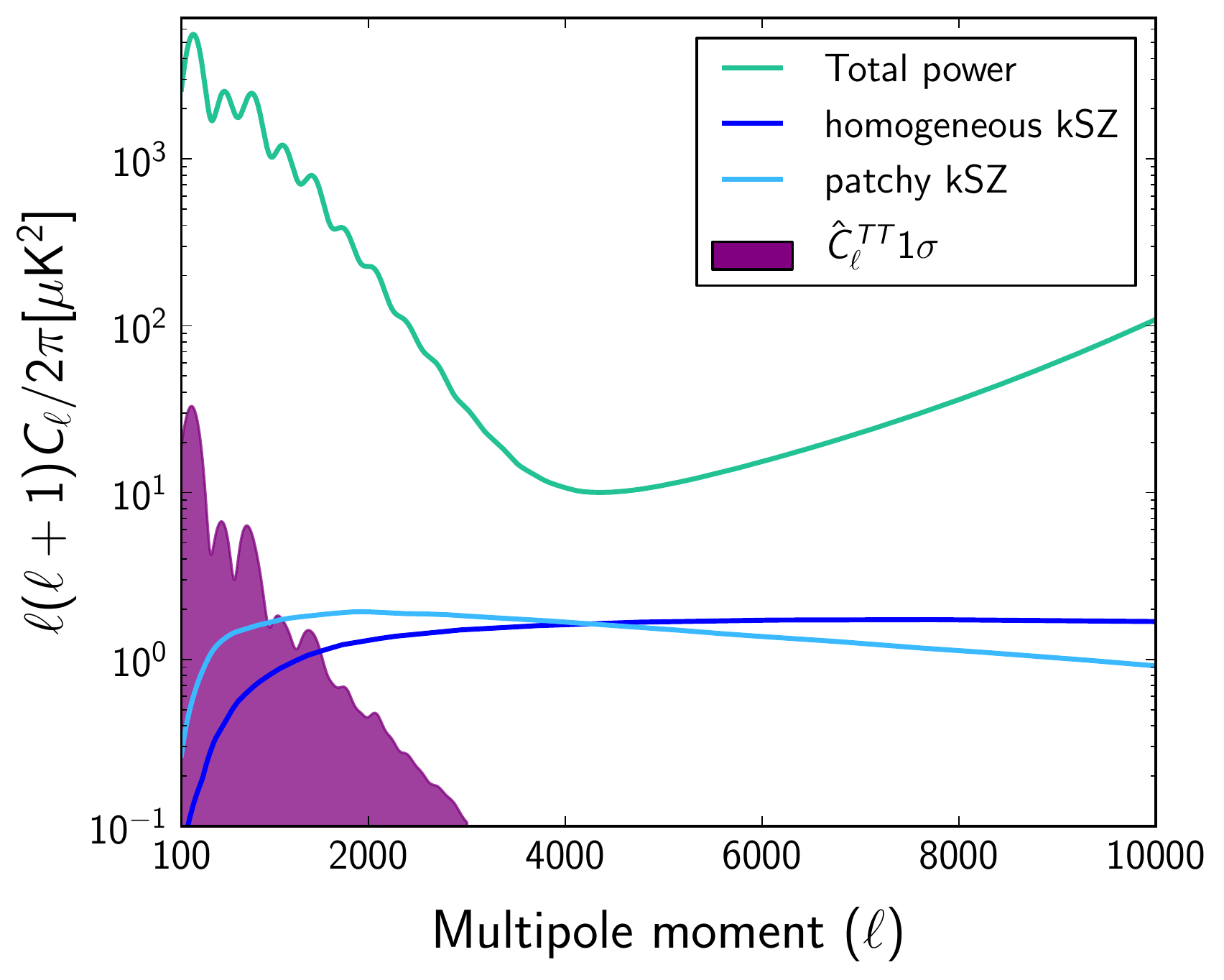}
\caption{The expected measured power at 150 GHz (green line) given by Eq.~\ref{eq:effvar} and computed from Planck 2013 {\em base\_planck\_lowl\_lowLike\_highL} cosmology, a combined kSZ signal contributing a total of $3.3\ \mu{\rm K}^2$ at $\ell=3000$, residual dust power at a level of $1\ \mu {\rm K}^2$ at $\ell=2500$, and map noise of 8 $\mu{\rm K}$-${\rm arcmin}$ after foreground cleaning for a beam resolution of $1.3^\prime$. It is compared to the $1\sigma$ uncertainty on the CMB temperature predicted by the combined Planck TT + AdvACT EE cosmology (purple band) and the theoretical kSZ signals (blue and cyan curves). The kSZ models are above the CMB uncertainty for $\ell>1500$ and, after subtracting the primordial CMB, are the main terms in the power in the region $2000<\ell<6000$, before noise starts to dominate.\\
\label{fig:ev}
}
\end{figure}
The actual kSZ constraining power of small-scale temperature data is shown in Fig.~\ref{fig:ev}, where we report the total power expected in 150 GHz CMB maps assuming most foregrounds have been cleaned out, computed as 
\begin{equation}
C^{\rm meas}_\ell = C^{\rm CMB}_{\ell}+C^{\rm kSZ}_{\ell}+C^{\rm Dust}_{\ell}+N_{\ell} \label{eq:effvar} \,,
\end{equation}
where the CMB contribution is the Planck 2013 {\em base\_planck\_lowl\_lowLike\_highL} cosmology. For the kSZ signal we account for homogeneous and patchy components contributing jointly with a combined amplitude of $1.5\ \mu {\rm K}^2+1.8\ \mu {\rm K}^2 = 3.3\ \mu{\rm K}^2$ at $\ell=3000$. We assume the cleaning using dusty high frequency channels leaves a residual Poisson dust power of $1\ \mu{\rm K}^2$ at $\ell=2500$ \citep{spergel2013}, and the noise level is obtained assuming a map noise of $8$ $\mu{\rm K}$-${\rm arcmin}$ for a beam resolution $\Theta_{\rm FWHM}=1.3^\prime$ after foreground cleaning.

Fixing the cosmology with small-scale polarization data, we in effect remove the primordial CMB contribution from temperature fluctuations. Then the high multipoles of the expected AdvACT TT data should provide a clean and definitive measurement of the kSZ signal, particularly in the minimum variance region $2000<\ell<6000$. At $\ell>1500$ the kSZ contributions are the main terms in the total power before noise starts to dominate.

\subsection{Measuring homogeneous kSZ}
\begin{figure}[t!]
\includegraphics[width=\columnwidth]{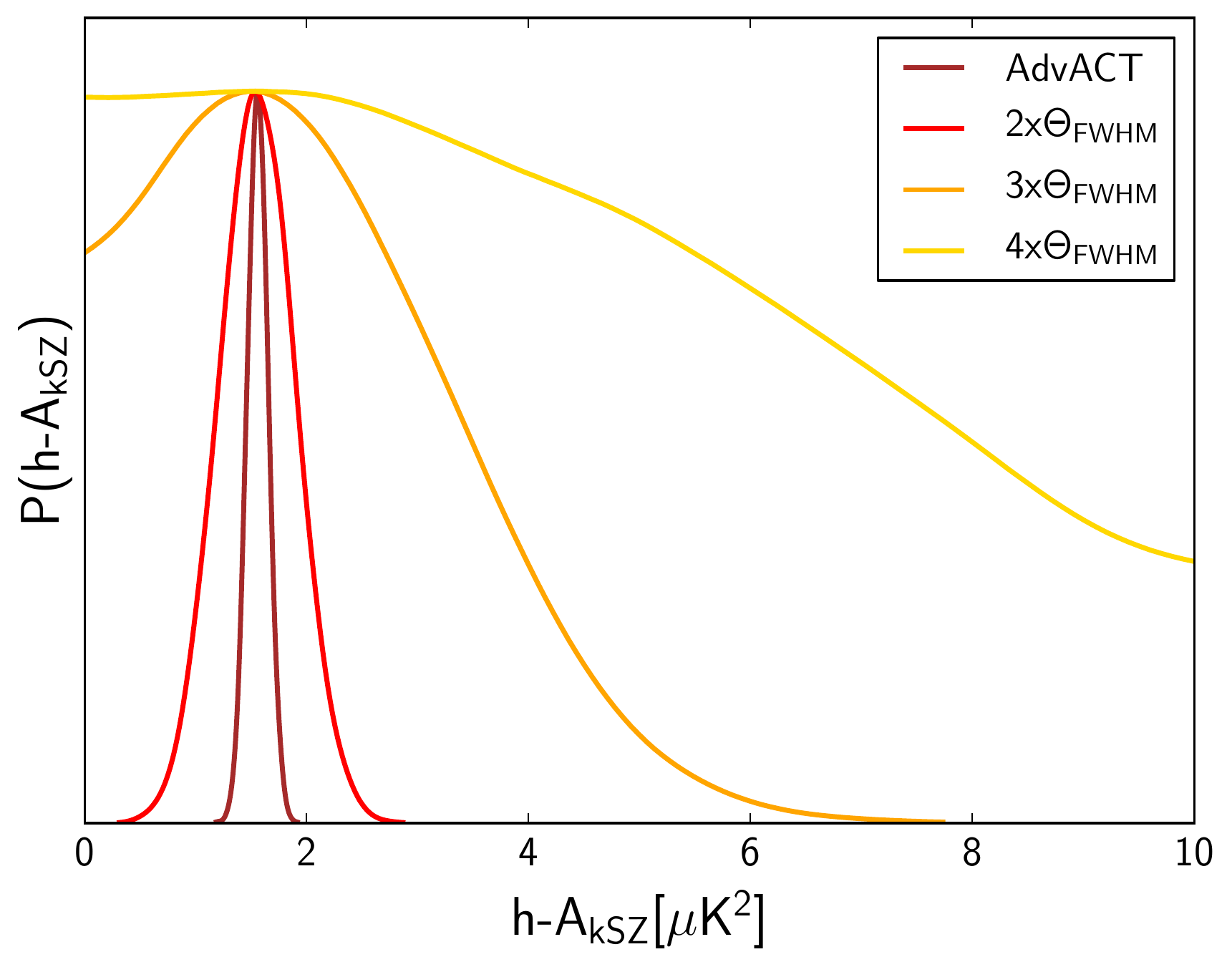}
\caption{Marginalized 1D posteriors of the homogeneous kSZ power at $\ell=3000$ plotted for an expected AdvACT beam, and for instruments with comparable sensitivity but 2, 3, and 4 times larger beam sizes. Larger beams dramatically reduce constraints on secondary foreground parameters.\\
\label{fig:kszb}
}
\end{figure}

To quantify this detection of the kSZ signal, we add the homogeneous kSZ signal to the simulated CMB signal, and an additional amplitude for the kSZ power at $\ell=3000$ to the fitted parameters in the MCMC analysis. We report the constraint on the homogeneous kSZ amplitude in Fig.~\ref{fig:kszb} and in particular we investigate the impact of the beam size on the detection. 
The constraint on the amplitude of the kSZ signal degrades considerably as one deviates from the projected beams of AdvACT, moving from a $15 \sigma$ detection with the assumed beams, $\sigma($h-A$_{\rm kSZ})$=0.1, to a completely unconstrained amplitude with beams that are 4 times larger than the baseline values.
Indeed, a larger beam will raise the tail of the total power significantly in the regime where the CMB TT signal is decaying, removing  the minimum variance window in multipole space where the kSZ can be constrained ($2000<\ell<6000$, see Fig.~\ref{fig:ev}) entirely.

Independence of the amplitude constraint on the cosmological model was tested by varying $N_{\rm eff}$: we found nearly identical limits.

The h-kSZ modelling is dependent on the underlying cosmology used in the simulations \citep{Shaw:2011sy}. Following the scaling relations and applying the cosmological corrections derived in \citet{Shaw:2011sy} we can translate the h-kSZ power at $\ell=3000$ into cosmological constraints. In particular, the amplitude of matter fluctuations on $8\ h^{-1} {\rm Mpc} $ scales, $\sigma_8$, varies the overall amplitude of the h-kSZ signal.
For the template used in this work 
\be
{\text {h-A$_{\rm kSZ}$}} \propto 1.65\ \Big (\frac{\sigma_8}{0.8}\Big)^{4.46} \mu {\rm K}^2 \, .
\ee
Analysing the simulated AdvACT data (including a h-kSZ component) results in a relative error on the matter fluctuations of $\sigma({\sigma_8})/\sigma_8= 0.01$. 

\subsection{Measuring patchy kSZ}
The patchy kSZ power spectrum amplitude and multipole shape depends on the time and duration of reionization. Limits on the reionization epoch from kSZ were first reported from SPT \citep{Zahn:2011vp} and ACT \citep{sievers2013} small-scale temperature data. 

If current theoretical models are correct, this reionization signal should be detectable at high significance with future ground-based CMB probes.
Removing the homogeneous contribution we can translate the p-kSZ amplitude, as we expect it to be measured by AdvACT with $\sigma($p-A$_{\rm kSZ})$=0.1, into a limit on the mean redshift and duration of reionization, $\zre$ and $\delz$ respectively, following \citet{Battaglia:2012im}
\be
{\text {p-A$_{\rm kSZ}$}}  = 2.03 \Big [ \Big ( \frac{1+{\zre}}{11} \Big )- 0.12  \Big ] \Big ( \frac{\delz}{1.05} \Big )^{0.51} \mu {\rm K}^2 \,.
\label{eq:pksz}
\ee

\noindent The definitions for the duration of reionization and redshift of 
reionization are $\delz \equiv z(\xe = 25\%) - z(\xe = 75\%)$ and
$\zre \equiv z(\xe = 50\%)$, respectively. Here $\xe$ is the ionization fraction of hydrogen. 
We note that Eq.~\ref{eq:pksz} is specific to the models in \citet{Battaglia:2012im} and is valid for $\delz > 0.2$; other models of reionization that produce large asymmetries in $\xe$ will deviate from this scaling relation \citep{Park2013}. 
Additionally, the shape of the patchy kSZ power spectrum is affected
by $\delz$ \citep{Zahn:2011vp,Mess:2012,Battaglia:2012im}. This precise
power spectrum shape is model dependent, but to first order it is a
function of the angular size of the ionized regions at $\zre$. In
general, for smaller $\delz$ values these ionized regions are larger,
thus, the power spectrum peaks on larger scales. The opposite follows
for larger $\delz$ values.
\begin{figure}[t!]
\includegraphics[width=\columnwidth]{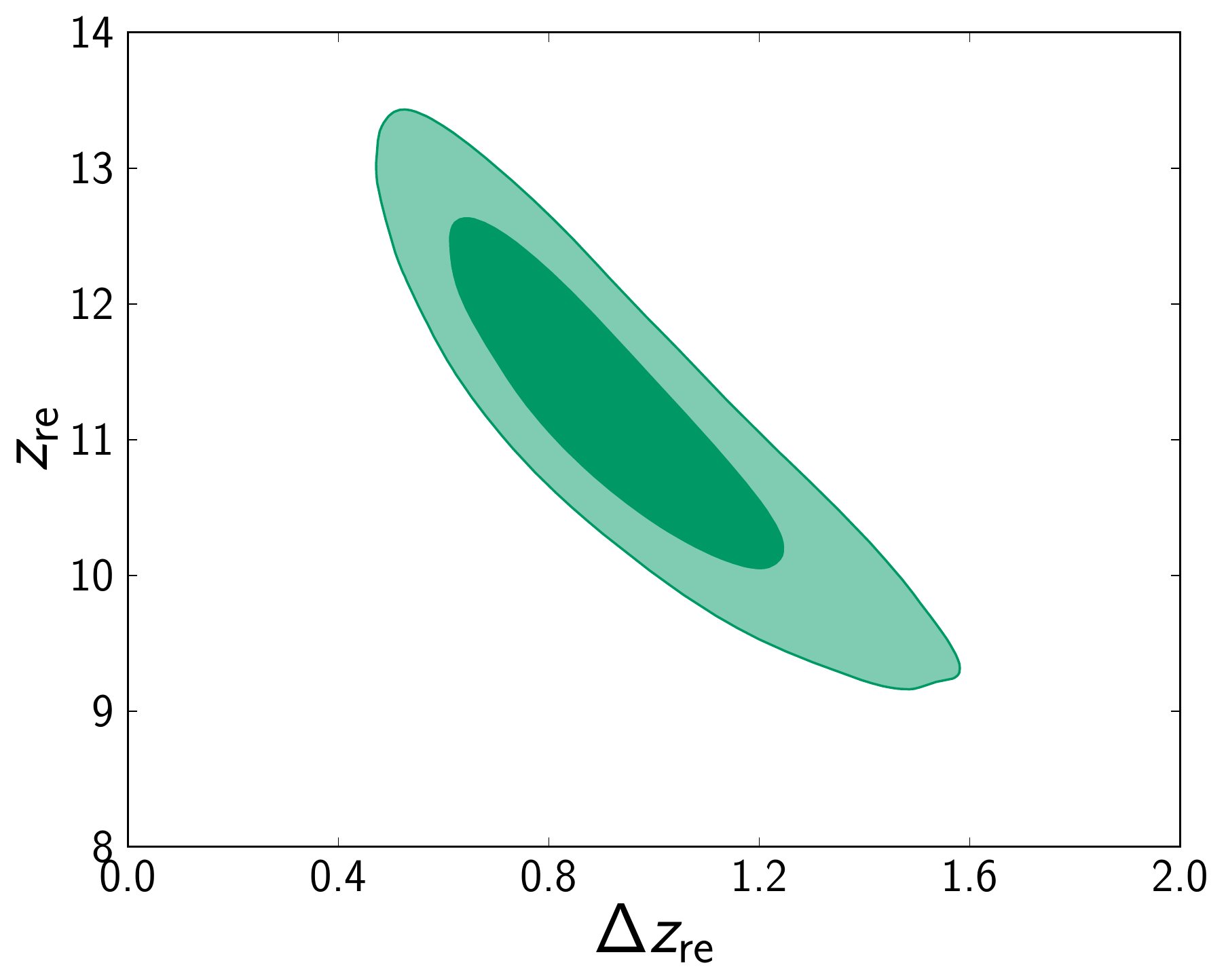}
\caption{Marginalized 2D contours at the 68 and 95\% confidence level in the $\delz - \zre$ plane as expected from AdvACT temperature and polarization observations in the range $300<\ell<3000$.\\
\label{fig:dzre}
}
\end{figure}

 Fig.~\ref{fig:dzre} shows the derived limits on the reionization parameters; the constraints on the optical depth from expected AdvACT data bound the reionization redshift with uncertainty $\sigma(\zre)=1.1$, while p-A$_{\rm kSZ}$ gives $\sigma(\delz)=0.2$. The figure is constructed by using Eq.~\ref{eq:pksz} and converting the optical depth, $\tau$, and p-A$_{\rm kSZ}$ into the quantities shown.
For a fixed amplitude of p-A$_{\rm kSZ}$, an extension of the duration of the reionization can be compensated with a decrement in the reionization redshift. Information on the optical depth breaks the degeneracy since $\tau$ is very sensitive to the mean redshift of reionization but not to its duration. 

A significant kSZ detection will allow us to place constraints on the physics of reionization, opening up the window on the universe at intermediate epochs.
This will be independent and complementary to the other reionization experiments such as those which are optimized to observe the hyperfine transition of neutral hydrogen (known as the 21 cm emission). In principal the measurements of the redshifted 21-cm power spectrum will provide tomographic measurements of reionization. However, due to the non-trivial complications of foreground removal, the current 21-cm experiments have placed upper limits on the power spectrum \citep[e.g.][]{Paciga2013,Parson2013} that are an order of magnitude higher than theoretical predictions. Thus, high-resolution polarized CMB experiments are well-positioned to improve on the constraints already in place from CMB temperature-only constraints \citep{Zahn:2011vp,sievers2013}.

\subsection{Going to lower multipoles} 
\begin{figure}[t!]
\includegraphics[width=\columnwidth]{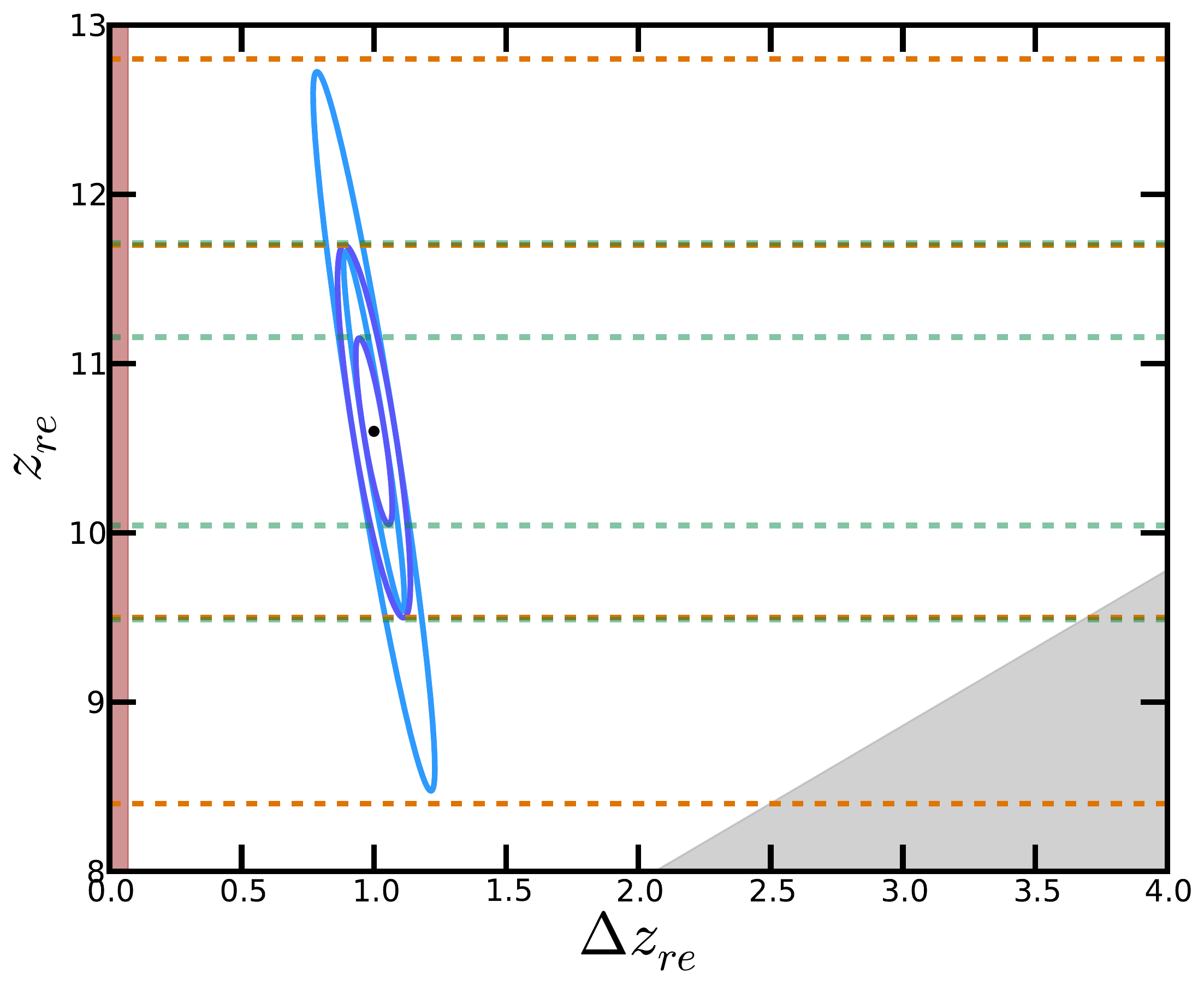}
\caption{2D contours at the 68 and 95\% confidence level in the $\delz - \zre$ plane as expected from AdvACT TT including patchy kSZ in combination with WMAP (light blue), or $10<\ell<3000$ AdvACT EE (dark blue) $\tau$ estimates. The orange and light green dashed lines report the 1 and 2$\sigma$ limits on $\zre$ from WMAP and AdvACT EE respectively. The brown and grey excluded regions are obtained from 21-centimetre line of atomic hydrogen and Gunn-Peterson Lyman-$\alpha$ absorption lines observations, respectively. \\
\label{fig:reio}
}
\end{figure}

The combination of small-scale temperature and polarization data as considered in this work constrains the optical depth with a $\sigma(\tau)=0.011$, comparable to current WMAP measurement ($\sigma(\tau)=0.013)$ from low-$\ell$ polarization.
If the proposed AdvACT measurements can be extended in multipole space down to $\ell=10$, we estimate an uncertainty in $\tau$ of $0.006$, although measuring such large scales may be challenging. 
This factor of 2 reduction in the uncertainty in $\tau$ informs our understanding of the epoch of reionization yielding $\sigma(\zre)=0.5$ and $\sigma(\delz) = 0.07$, as shown in Fig.~\ref{fig:reio}. We report in Fig.~\ref{fig:reio} Fisher matrix calculations propagating the shape and amplitude of the patchy kSZ model in combination with priors on the optical depth. The light blue contours report the region bounded by a $\tau$ estimate from WMAP (i.e. assuming $\sigma(\tau)=0.013$) in combination with AdvACT TT patchy kSZ, while the dark blue region shows the better limits that would result from measuring $\tau$ using AdvACT EE data at $10<\ell<3000$. The dashed curves project the 1 and 2$\sigma$ errors on $\zre$ from WMAP (orange) and AdvACT (light green). We compare with other astrophysical limits reporting the excluded regions by the 21-centimetre line of atomic hydrogen \citep{Bowman:2012hf} (brown band) and Gunn-Peterson Lyman-$\alpha$ absorption lines \citep{Fan:2006dp} (grey band).

\section{Discussion}
\label{sec:disc}
The upcoming measurements of CMB polarization will provide both an independent confirmation of the gravitational instability picture for structure formation and a useful consistency check on the temperature observations; it will also offer constraining power for the cosmological model. 
Future small-scale CMB experiments will build on the successes of the current suite of ground-based CMB probes such as ACT to provide high resolution and sensitivity measurements of the small-scale CMB polarization over a wide frequency and sky coverage. In this paper we illustrated the power of future arcminute resolution polarization data, using as an example expected data from a possible extension of the ACTPol project, AdvACT, in predicting the contribution to the high-$\ell$ tail of the CMB temperature spectrum from primordial anisotropies. In addition, combining future AdvACT EE data with expected full mission Planck temperature data yields tight constraints on the cosmological model, bounding the damping tail of the CMB to exquisite precision. 

The expected precision with which a future experiment like AdvACT will measure the EE polarization spectrum will usher in a new epoch for CMB cosmology. By estimating the primordial cosmology with small-scale polarization measurements, one can also use the measured small-scale temperature to probe the wide range of astrophysical processes affecting the temperature spectrum in the high multipole region directly. The broad frequency coverage provided by Planck observations and future AdvACT data will separate out most of the secondary galactic and extra-galactic emissions contributing to the temperature data. The bound on the primordial power coming from polarization data will then unveil residual emissions. 
In this work we have assumed an efficient cleaning of all foregrounds except the only secondary component degenerate in frequency space with the lensed primary fluctuations and hence left after cleaning, namely the kSZ term. We note that a non-perfect cleaning will increase the uncertainty of the kSZ signal measurement, however we have shown that a window in the multiple space region $2000<\ell<6000$ where the kSZ dominates the total power will provide a definitive measurement of this signal.

We estimate a $15 \sigma$ detection of the diffuse, homogeneous kSZ signal from AdvACT small-scale temperature data leading to a measurement of the amplitude of matter density fluctuations, $\sigma_8$, at $1\%$ precision.
Alternatively, using the reionization signal encoded in the patchy kSZ measurements we also predict bounds on the time and duration of reionization, with uncertainty $\sigma(\zre)=1.1$ and $\sigma(\delz)=0.2$ respectively. We find that these kSZ detections are strongly dependent on sensitivity and beam size, highlighting the importance of future CMB surveys with arcminute-scale resolution.

\acknowledgements
We acknowledge Graeme Addison and Sigurd Naess for useful discussions.
This work was supported by the U.S. National Science Foundation through awards
AST-0408698 and AST-0965625 for the ACT project, as well as awards PHY-0855887
and PHY-1214379. Funding was also provided by Princeton University, the
University of Pennsylvania, and a Canada Foundation for Innovation (CFI) award
to UBC. ACT operates in the Parque Astron\'omico Atacama in northern Chile
under the auspices of the Comisi\'on Nacional de Investigaci\'on Cient\'ifica y
Tecnol\'ogica de Chile (CONICYT). 
Funding from ERC grant 259505 supports EC and TL. 
HT is supported by grants NASA ATP NNX14AB57G and NSF AST- 1312991.
EC thanks Princeton Astrophysics for hospitality during this work. 
 

\end{document}